\documentclass[aps,prl, amsfonts, amssymb, amsmath, reprint, eprint, showkeys, nofootinbib, twoside,    longbibliography]{revtex4-2}
\usepackage{amsmath,amssymb,amsfonts}

\usepackage[english]{babel}
\usepackage[utf8]{inputenc}

\usepackage[colorinlistoftodos, color=green!40, prependcaption]{todonotes}

\usepackage{graphicx}
\usepackage{dcolumn}
\usepackage{amsmath,amsbsy,amssymb,scalefnt}
\usepackage{float} 
\usepackage{lipsum}
\usepackage{verbatim}
\usepackage{bm}
\usepackage{epstopdf}
\usepackage{amsfonts}
\usepackage{textcomp}
\usepackage{soul}
\usepackage{mathtools}
\usepackage[colorlinks=true]{hyperref}
\hypersetup{
  bookmarksnumbered,
  pdfstartview={FitH},
  citecolor=black,
  linkcolor=red,
  urlcolor=black,
  pdfpagemode=UseOutlines
}
\usepackage{comment}
\usepackage{fancyhdr}
\usepackage{mathtools}
\usepackage[normalem]{ulem}
\usepackage[x11names]{xcolor}

\usepackage{amsthm}
\usepackage{mathtools}
\usepackage{physics}
\usepackage{xcolor}
\usepackage{graphicx}
\usepackage[left=23mm,right=13mm,top=35mm,columnsep=15pt]{geometry} 
\usepackage{adjustbox}
\usepackage{placeins}
\usepackage[T1]{fontenc}
\usepackage{lipsum}
\usepackage{csquotes}
\usepackage[colorlinks=true]{hyperref}

\usepackage{comment}

\usepackage{fancyhdr}
\usepackage[normalem]{ulem}
\usepackage[x11names]{xcolor}

\includecomment{taglio}

\begin{document}
\pagestyle{plain}
\title{An Information-Theoretic Bound on Thermodynamic Efficiency \\ and the Generalized Carnot's Theorem}

\author{Anna Gabetti}
\email[Correspondence email address: ]{anna.gabetti@polito.it}
\author{Fabrizio Dolcini}
\author{Davide Girolami}

    \affiliation{Dipartimento di Scienza Applicata e Tecnologia, Politecnico di Torino, Corso Duca degli Abruzzi 24, Torino, Italy
}

\begin{abstract}
We derive a bound on the efficiency of thermal engines that can be sharper than Carnot's limit. It is a function of statistical correlations between the engine internal state and Hamiltonian, can be saturated even in finite-time cycles, and applies to both classical and quantum engines. Specifically, the bound establishes the exact maximal efficiency of engines operating with multiple baths, tightening the upper limit set by Carnot's theorem. 
Then, we show that an engine made of a quantum dot coupled with fermionic baths can achieve the bound, even when operating beyond the quasistatic regime. The bound relates thermodynamic efficiency to statistical correlations of the working medium and identifies conditions for its saturation in classical and quantum cycles.
\end{abstract}

\maketitle

\section*{Introduction} Optimizing the performance of engines is  a central challenge of modern technology, which requires maximizing work extraction, while minimizing  energy dissipation. Specifically, the simplest thermal machine  absorbs heat $Q_h>0$ from a hot reservoir (bath) at temperature $T_h$ and converts part of it into work $W>0$, while dumping the residual heat $Q_c = W-Q_h<0$  into a cold bath at temperature $T_c$.
The second law of thermodynamics sets a general upper bound on the efficiency $\eta$ of heat-to-work conversion in cycles operating between {\it two} thermal baths, 
stated by Carnot’s theorem \cite{callen,carnot}:
\begin{align}\label{eq1}
    \eta:=1-\frac{|Q_c|}{Q_h}\leq \eta_C,\qquad\eta_C=1-\frac{T_c}{T_h}.
\end{align}
However, its practical relevance is limited, for the bound is only reached  in  reversible cycles. This idealized scenario requires a quasi-static (infinite-time) heat exchange with the baths, while preserving their sharply defined temperatures. Real thermal machines instead interact for a finite time with baths through  temperature gradients or with multiple baths at different temperatures (e.g., Otto and Diesel cycles). Furthermore, $\eta_C$ provides no guideline for optimizing the use of the engine internal resources, since it only depends on environmental properties (i.e., bath temperatures).

Here, we derive a bound on thermodynamic efficiency, applicable to both classical and quantum systems, which depends on controllable physical parameters of an engine. Specifically, it is a function of how much we know about its state, how fast such information changes, and the accessible energy levels, determined by the engine Hamiltonian. Since it depends on the engine internal dynamics, this bound inherently differs from the known refinements of Carnot's theorem, which are limited to the two-bath scenario and identify  an engine maximal efficiency exclusively in terms of bath temperatures  \cite{carnot1,carnot2,carnot3,carnot4,carnot5,carnot6,carnot7,carnotgen,EspositoCarnot,EspositoMaxPower,EspositoFinitetimeDot,carnot8,carnot9}. \\
Crucially,  the bound can be much sharper  than Eq.~(\ref{eq1}) for non-reversible cycles and for reversible ones involving multiple baths, while exactly matching $\eta_C$ for reversible cycles involving two baths.
Remarkably, the bound can be saturated in finite-time classical processes with imperfect control of the system driving. Quantum coherence in the energy eigenbasis is therefore
not necessary to saturate the information-theoretic bound. Its effect on the actual efficiency depends
on the engine dynamics and the admissible controls.\\
We design a single-level quantum dot coupled to two fermionic baths at fixed temperatures and chemical potentials. This model offers an example of a  finite-time cycle that saturates  the
information-theoretic bound. We also investigate
how stochastic fluctuations in the control of its
energy level affect the efficiency relative to
the noise-free reference protocol. This cycle  is experimentally implementable with current technology \cite{dot1,dot2,dot3,dot4,dot5,dot6,pekola}.\\
We also evaluate the performance of a classical engine with two fermionic channels, i.e. {\it four} occupancy states, and of a two-level system, whose quantum state also displays coherence in the energy eigenbasis. For the protocols and parameters considered, we find that  additional energy levels and coherence degrade, rather than enhance, the engine  performance. The efficiency only achieves the upper efficiency bound when the engine approaches the two-level classical case.\\

\section{Derivation of the bound} Consider a quantum system (the engine under scrutiny) in the state $\rho(t)$ and with Hamiltonian  $H(t)$ that exchanges energy with its environment.
In the Markovian regime, the system evolution is governed by the Lindblad equation:
\begin{align}\label{lindblad}
\dot \rho(t) = -i [H(t),\rho(t)]+ {\cal L}_t(\rho(t)) \quad,
\end{align}
where ${\cal L}_t(\rho(t))$ describes the interactions with the environment \cite{cmp/1103899849,10.1093/acprof:oso/9780199213900.001.0001}, including heat exchanges with baths. We here set $\hbar=1$. The rate of change of the system internal energy is $\dot U(t)=d\text{Tr}\{\rho(t)H(t)\}/dt$, and the heat current is $\dot{Q}(t):=\text{Tr}\left\{\dot{\rho}(t)H(t)\right\}$ \cite{Alicki,qthermorev}. The efficiency of thermodynamic cycles, $\int \dot U(t) dt =0$, is given by 
\begin{align}\label{eff}
\eta=1-\frac{\int_{Q_-}|\dot Q(t)| dt}{\int_{Q_+} \dot Q(t) dt},
\end{align}
where the integrals $\int_{Q_{-(+)}}$ are computed over the heat-release (absorption) portions of the cycle.\\
Our first main result is:\\
{\bf Result 1 --} {\it  The efficiency of a thermodynamic engine
satisfies
\begin{align}
\eta\leq \eta_{max}=1-\frac{\displaystyle\int_{Q_-} -\mathcal{I}_{max}(t)\, dt}{\displaystyle\int_{Q_+}\;\mathcal{I}_{max}(t) \,dt},\label{bound1}
\end{align}
where the maximal heat current $\mathcal{I}_{max}(t)$  is a quantity that depends only on $\rho(t)$, $\dot\rho(t)$ and $H(t)$.}
\begin{proof} We express the heat current $\dot Q(t)$ in terms of the linear correlation between the evolution rate of the system state and its Hamiltonian \cite{Green,Garc_a_Pintos_2022}. We recall the definition of  symmetrized covariance $ \text{Cov}(X,Y):=\,  \text{Tr}\{\left\{X,Y\}\rho(t)\right\}/2-\text{Tr}\{X \rho(t)\}\text{Tr}\{Y \rho(t)\}$ for any  two Hermitian operators $X$ and $Y$. Then, one has
\begin{equation}\label{Qdot-corr}
\dot{Q}(t)= \, \text{Cov}\left(L(t),H(t)\right) \quad,
\end{equation}
where $L(t)$ is the symmetric logarithmic derivative \cite{quantumgeometry}, defined  by $\dot{\rho}(t) = (L(t)\rho(t) + \rho(t)L(t))/2$, which reduces to $L=\rho^{-1}\dot\rho=d\ln\rho(t)/dt$ in the classical case
$[\rho(t),\dot\rho(t)]=0$.\\
The covariance is maximal (in absolute value) for linearly dependent quantities \cite{cover}, while there is no heat current in adiabatic transformations.\\
Next, the covariance between two variables can be bounded in terms of their covariances with respect to a third variable, which we choose to be $\ln(\rho(t))$. Then,
\begin{align}
\dot{S}(t)=& -k_B \text{Tr}\left\{\dot{\rho}(t)\ln\rho(t)\right\}=-k_B\text{Cov}\left(L(t),\ln\rho(t)\right)\nonumber\\
{G}(t):=\, & \text{Cov}\left(H(t),\ln\rho(t)\right).
\end{align}
Here, $\dot{S}$ is the  rate at which the system state varies its thermodynamic entropy, while  $G(t)$ reaches maximal absolute values for Gibbs states $\rho(t)\propto e^{-\beta(t)\,H(t)}$.\\
It is then possible to construct the correlation matrix $\text{corr}(V(t))$ of the  triad of observables 
$V(t)=\left\{L(t), \ln \rho(t), H(t)\right\}$. Since it is real and positive semidefinite, it satisfies $\text{det}\left(\text{corr}(V(t))\right)\geq 0$, which implies $\mathcal{I}_{min}\leq\dot Q(t) \leq \mathcal{I}_{max}(t)$, where
\begin{align}
&\mathcal{I}_{max(min)}(t) := \nonumber\\
&\left( R_S(t) R_G(t) +(-) \sqrt{(1-R_S(t)^2)(1-R_ G(t)^2)}\right)\nonumber\\
&\text{SD}(H(t))\ \text{SD}\left({L(t)}\right),\nonumber\\
&R_S(t):=\frac{-\dot S(t)/k_B}{\text{SD}(L(t))\,\text{SD}(\ln\rho(t))},\nonumber\\
&R_{G}(t):=\frac{G(t)}{\text{SD}(\ln \rho(t))\,\text{SD}(H(t))}.
\end{align}

Here, $R_{S,G}(t)\in[-1,1]$ are the symmetrized Pearson correlation coefficients of $-\dot{S}(t),\,G(t)$, and $\text{SD}(X)=\sqrt{\text{Tr}\{\rho(t)X^2\}-(\text{Tr}\{\rho(t)X\})^2}$ is the standard deviation of~$X$. Hence, during the time intervals over which heat is absorbed by the engine, $\dot Q(t)>0$, one has
$\int_{Q_+}\dot Q(t)\,dt\le\int_{Q_+}\mathcal I_{max}(t)\,dt$.
By contrast, when  heat is released, $\dot Q(t)<0$, one has
$\int_{Q_-}|\dot Q(t)|\,dt
=-\int_{Q_-}\dot Q(t)\,dt
\ge-\int_{Q_-}\mathcal I_{max}(t)\,dt$.
These inequalities imply the upper bound to the efficiency,   Eq.~(\ref{bound1}).
\end{proof}
 
The result limits heat exchanges in terms of  statistical correlations among (functions of) a system's internal properties (state and Hamiltonian). The derivation does not require a specified reservoir model,
although the values of the bounds depend on the environment
through the system dynamics. The conditions for saturating this bound concern the
statistical properties of a given cycle. Establishing
optimality within a family of admissible protocols
additionally requires specifying the shared physical
constraints, since varying the controls generally
changes both the trajectory and the bound.\\
We note that the  inequality $\dot Q(t)\geq \mathcal I_{min}(t), \forall t$, implies an information-theoretic {\it lower} bound to the engine's efficiency, provided that
$\int_{Q_+}\mathcal I_{\min}(t)\,dt>0$:
\begin{equation}\label{lower}\eta_{min}
:=
1-
\frac{\displaystyle\int_{Q_-} -\mathcal I_{min}(t)\,dt}
{\displaystyle\int_{Q_+}\mathcal I_{min}(t)\,dt}
\leq\eta \quad.
\end{equation}
Hence, when $\det(\mathrm{corr}(V(t)))=0$, the heat
current saturates at least one of its local bounds $\mathcal I_{max}(t),\mathcal I_{min}(t)$.
The efficiency saturates the corresponding bound
in Eqs.~(\ref{bound1}) and (\ref{lower}) provided
the same local branch is followed almost everywhere
on both the heat-absorption and heat-release portions
of the cycle. If $|R_S(t)|=1$ or $|R_G(t)|=1$ $\forall t$, the two bounds coincide, $\eta_{min}=\eta=\eta_{max}$.
This occurs for Gibbs states $\rho(t)=e^{-\beta(t)\, H(t)}/Z(t),\,\forall t$,  where $\beta(t)$ is the instantaneous inverse temperature of the system, and $Z(t)$ is the partition function. Note that the engine does not need to be in equilibrium with the environment: the bound is saturated even if $\beta(t) \neq 1/k_B T_{\text{env}}(t)$. That is, it can be achieved even in irreversible cycles.\\
 Yet, there exists a relation between the new  bound and the standard formulation of the second law of thermodynamics in Eq.~(\ref{eq1}). When the engine reaches thermal equilibrium with the environment, the information content of its state explicitly depends on the environment temperature.
 Indeed, Eq.~($\ref{bound1}$) generalizes Carnot's result to the multiple-bath scenario, setting an informative upper limit to the efficiency of realistic thermal engines:\\

{\bf Result 2 (Generalized Carnot's Theorem) --} {\it For cyclic transformations of thermal machines, one has:
\begin{align}\label{gencarnot}
\eta\leq \eta^{\cal R}_{max}=  1-\frac{\overline{T}_-}{\overline{T}_+},
\end{align}
where $\overline{T}_-$ and $\overline{T}_+$ are the entropy-weighted average temperatures of the environment,
\begin{align}
\overline{T}_{-(+)}
&:=\frac{\int_{S_{-(+)}} T_{\mathrm{env}}(S)|dS|}{\int_{S_{-(+)}} |dS|} \quad,
\end{align} \noindent computed over the entropy-decrease ($S_-$) and entropy-increase ($S_+$) regions, respectively. The bound is saturated ($\eta^{{\cal R}} =\eta^{\cal R}_{max}$) for reversible cycles.}

\begin{proof}
Consider two thermal machines $\cal{R}$ and $\cal{X}$ that run a reversible cycle and a generic one, respectively, in contact with the same environment.
They exchange heat with an arbitrary number of different baths, whose instantaneous temperature is denoted by $T_{\text{env}}(t)$.\\
The engine $\cal{R}$  is at  thermal equilibrium with the environment at all times: $\ln \rho^{\cal{R}}(t) = - H^{\cal{R}}(t)/k_B T^{\cal{R}}(t)-\ln Z^{{\cal R}}(t)$ ($T_{\text{env}}(t)=T^{\cal{R}}(t)$). 
As a result, ${\cal I}_{max}^{\cal{R}}(t)= \dot{S}^{{\cal R}}(t)   T^{{\cal R }}(t), \,\forall\, t$. Further, in non-adiabatic regions of a reversible cycle, $\dot{S}^{{\cal R}}(t)T^{{\cal R }}(t)\,dt=dS^{{\cal R}}\,T^{\cal R}(S^{{\cal R}})$. For  cyclic transformations, one has
$\int_{S_-}|dS^{\cal R}|=\int_{S_+}dS^{\cal R}$. At the same time, the  Clausius inequality is saturated: $\text{d}S^{\cal{R}}(t)= \frac{\delta Q^{\cal{R}}(t)}{T_{\text{env}}(t)}$ \cite{clausius}. Hence,  
\begin{align}
\eta^{\cal{R}}=\eta_{max}^{\cal{R}} =1-\frac{\displaystyle\int_{S_-} |dS^{{\cal R}}|\,T^{\cal R}(S^{{\cal R}})} {\displaystyle\int_{S_+}\; dS^{{\cal R}}\,T^{\cal R}(S^{{\cal R}})}=1-\frac{\overline{T}_-}{\overline{T}_+}.\end{align} 
Note that the temperature is averaged with respect to entropy, which is generally different from the temporal average, unless the entropy changes linearly in time.
For the generic machine ${\cal X}$, one has $Q_{h}^{\cal X}\leq \int_{S_+} T_{\text{env}}(S^{{\cal X}})d S^{\cal X}$ and $| Q_{c}^{\cal X}|\geq\int_{S_-}  T_{\text{env}}(S^{{\cal X}})|d S^{{\cal X}}|$, which are saturated  if and only if the cycle is reversible. Hence, the claim is proven, as $\eta^{\cal X}\leq \eta^{{\cal R}}$.\\
\end{proof} 

Here we emphasize that the inequality in Eq.~(\ref{gencarnot}) is generally much tighter than the extremal-temperature Carnot bound $1-T_{\text{min}}/T_{\text{max}}$, where $T_{\text{min,max}}$ are the minimal/maximal temperatures at which heat is released/absorbed \cite{carnotgen}. The latter
is saturated only by reversible cycles exchanging heat
exclusively with the two baths at $T_{\min}$ and
$T_{\max}$, whereas our expression is attained by
reversible cycles involving an arbitrary number of
baths, with the entropy-weighted temperatures evaluated
along the corresponding cycle. Also, while an  irreversible cycle $\cal X$ can have $\eta^{{\cal X}} < \eta^{\cal R}_{max}< \eta_{max}^{{\cal X}}$, it is possible that $\eta^{{\cal X}}\leq\eta_{max}^{{\cal X}}<\eta_{max}^{{\cal R}}$. That is,  the upper limit in Eq.~(\ref{bound1}) can be more stringent than Carnot's. In particular, the following example illustrates that the
information-theoretic bounds can be saturated in
a finite-time, irreversible cycle, $\eta^{{\cal X}}=\eta_{max}^{{\cal X}}$. We consider
a diagonal two-level engine as a simple explicit
realization of this property.\\

\section{Case Study} 
\begin{figure}[t!]  \centering\includegraphics[width=1\linewidth]
{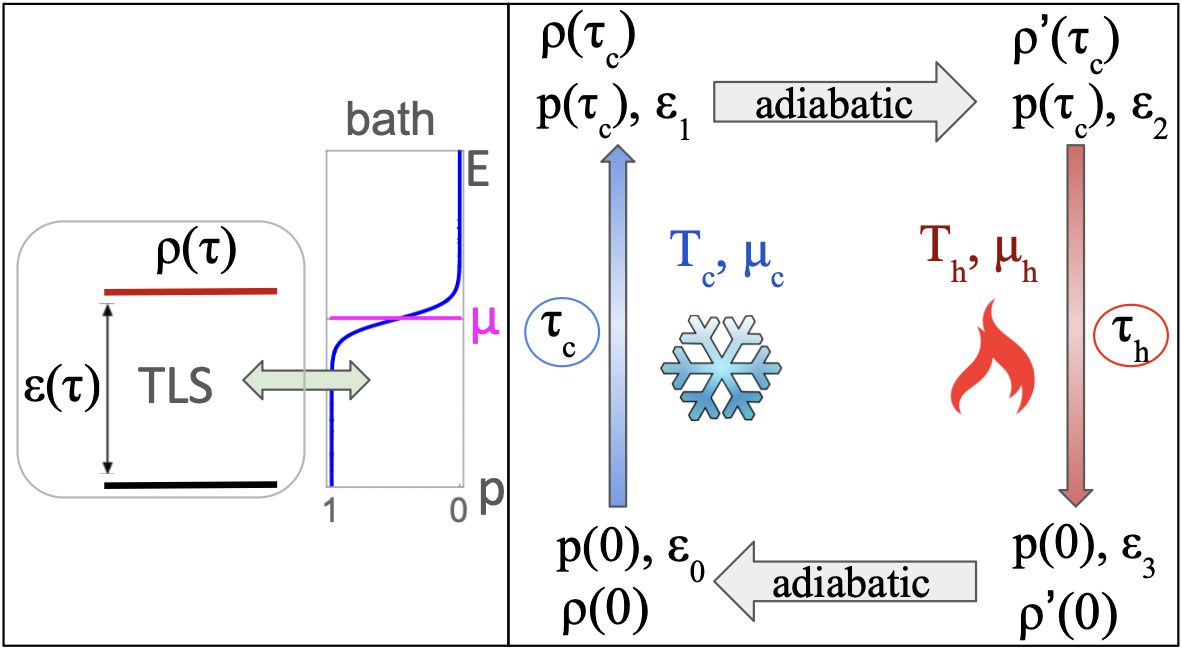} 
\caption{We study the efficiency of a two-level system (TLS) that undergoes a finite-time cycle consisting of two driven thermal contacts and two adiabatic transformations, which mirrors the Carnot cycle in the quasi-static limit.}
   \label{fig2}
\end{figure}

Let us consider   a single-level quantum dot coupled to a metallic electrode (fermionic bath)  at temperature $T$. The electron hopping from/to the electrode corresponds to an energy exchange between the system and the bath, which is determined by the deviation $\epsilon$ of the  dot energy level from the bath chemical potential $\mu$. Moreover, $\epsilon=\epsilon(t)$ is externally controlled in time through a gate voltage applied to the dot. The dot  occupancy levels (empty and filled) are represented by the ground and excited states $|\downarrow\rangle,|\uparrow\rangle$ of a two-level system Pauli operator $\sigma_z$, respectively. Then, the dot     Hamiltonian is $H(t)= \epsilon(t)  (\mathbb{I}+\sigma_z)/2$, and its  state $\rho(t)$ is a diagonal density matrix, fully characterized by the populations 
$p(t)$ and $1-p(t)$, where $p(t)$ is the probability that the dot level is occupied at time $t$. This choice does not imply loss of generality: noncommutativity between engine state and Hamiltonian   cannot help break  Carnot's limit \cite{carnotquantum,ThermoEngine}, and we here show that this classical engine already achieves the bound in Eq.~(\ref{bound1}).\\
The system undergoes a cycle of four transformations, as follows (see Figure~\ref{fig2}).\\
(i) Initially, the system is at equilibrium with the cold  bath at temperature $T_c$ and chemical potential $\mu_c$.
 Within a time $\tau_c$, the energy level increases according to the  linear protocol  $\epsilon_c(t)= \epsilon_* f_c(t)=\epsilon_0+2t(\epsilon_1-\epsilon_0)$. Here, $\epsilon_*$ denotes the characteristic energy scale of the system, with $t\in[0, 1/2]$ being the dimensionless time variable, i.e. the   physical time in units of $\tau_c+\tau_h$. The transformation results in work performed onto the system, which releases heat. Only in the quasistatic limit $\tau_c \rightarrow \infty$, the system remains arbitrarily close to the instantaneous equilibrium with the bath, and the transformation becomes reversible, corresponding to an actual isothermal process. We set $\epsilon_0\equiv\epsilon_c(0)$ and $ \epsilon_1\equiv\epsilon_c(1/2)$ to be the initial and final energy values of the transformation. \\(ii) The second stroke is adiabatic. The system evolves unitarily, and the energy level  shifts from $\epsilon_1$ to $\epsilon_2$. Since the populations remain constant, no heat is exchanged. The transformation duration is assumed negligible compared to the timescale of the whole cycle.\\ 
(iii) Next, the dot is coupled to the hot bath at temperature $T_h$ and chemical potential $\mu_h$. Within a time $\tau_h=\tau_c$, the energy level follows the linear protocol $\epsilon_h(t)=\epsilon_* f_h(t)=\epsilon_2-v(t-1/2)$ over $t \in (\frac{1}{2},1)$, decreasing from $\epsilon_2=\epsilon_h(1/2)$ to $\epsilon_3=\epsilon_h(1)$, allowing the system to absorb heat and to perform work. The parameter~$v$ specifies the protocol slope. To avoid a discontinuity in the equilibrium populations at the switching point $t=1/2$, which would lead to an instantaneous relaxation and additional entropy production, we impose the continuity condition on equilibrium states, $\beta_c\epsilon_c(1/2)=\beta_h\epsilon_h(1/2)$, fixing $\epsilon_2.$ In the reversible limit, cycle closure imposes an additional constraint on the final energy, $\beta_c\epsilon_c(0)=\beta_h\epsilon_h(1)$,  fixing $\epsilon_3$ and the protocol slope $v$. These two conditions together set the energy variations in the reversible limit to satisfy $\frac{\Delta\epsilon_h}{\Delta\epsilon_c}=\frac{T_h}{T_c}.$ At finite time, dissipative effects prevent cycle closure with fixed parameters. Thus, the slope $v$ is fixed to ensure the cyclic condition $p(0)=p(1)$.\\
(iv) The cycle closes with a second, very fast adiabatic transformation, bringing the energy level from $\epsilon_3$ back to $\epsilon_0$, with an associated work.
The system is finally reconnected to the cold bath, completing the cycle. The process has a total duration $2\,\tau_c=2\,\tau_h$, reducing to a reversible Carnot cycle in the quasi-static limit $\tau_h, \tau_c \to \infty$.\\
The dynamics of the excited level population $p(t)$ during the two strokes in which the system is coupled to a bath at inverse temperature $\beta$ is governed by the master equation \cite{qthermorev,Alicki,10.1093/acprof:oso/9780199213900.001.0001,cmp/1103899849,potts}
\begin{align}
      \dot{p}(t)=\gamma_\uparrow\, (1-p(t))-\gamma_\downarrow \,p(t)\label{mastereq} \quad.
\end{align} 
The transition rates are $\gamma_\uparrow (t)=c /(1 + e^{\beta \epsilon(t)})$ and $ \gamma_\downarrow(t)=c /(1 + e^{-\beta \epsilon(t)})$, corresponding to tunneling into and out of the dot, respectively. They are derived within the wide-band approximation and satisfy quantum detailed balance \cite{EspositoFinitetimeDot}.
Here, $c=\Gamma\tau$ is the dimensionless dot-bath coupling parameter, with $\Gamma$ the tunneling rate and $\tau$ the transformation duration. The population relaxes toward thermal equilibrium, with deviations decaying exponentially at
rate $\gamma_\downarrow+\gamma_\uparrow=c$.\\
Initially, the system is in equilibrium with the cold bath at inverse temperature $\beta_c$,
$\rho(0) = Z^{-1} \exp[-\beta_c H(0)] = 
{\rm diag}[ p(0), 1 - p(0)]$, with $p(0) = 1/(1 + e^{\beta_c \epsilon(0)})$. In the absence of coupling ($c=0$), the state evolves unitarily, i.e., $ p(t)$ stays fixed at its initial thermal equilibrium state $p(0)$. In the quasistatic limit $c\gg1$, i.e., when the cycle duration $\tau_c+\tau_h$ is much longer than the relaxation time of the system $\Gamma^{-1}$, the system remains arbitrarily close to the instantaneous Gibbs state of its Hamiltonian,
realizing an ideal isothermal process. 
\begin{figure}[t!]
    \centering
\includegraphics[width=0.48 \textwidth]{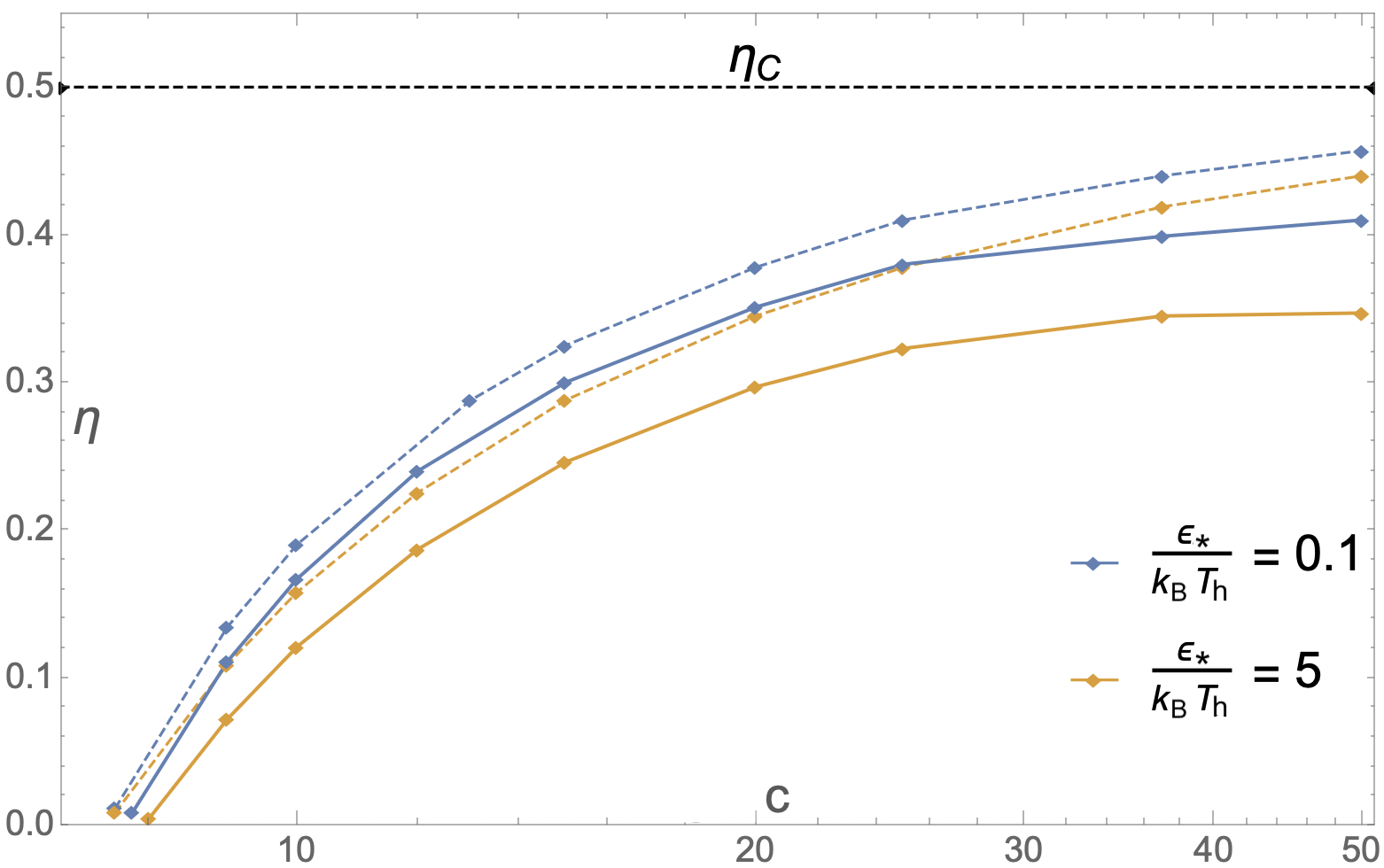}
\caption{\label{fig:bounds}
The cycle efficiency $\eta$ (solid line) and its bound
$\eta_{max}$ (dashed line) are shown as a function of the dimensionless system-bath coupling strength $c$, for two different values of  $ \epsilon_*/k_B T_h$. The efficiency increases monotonically with $c$, approaching $\eta_{\text{C}}=0.5$ in the reversible limit $c \rightarrow \infty$. Notably, $\eta_{max}$  exhibits the same increasing trend as $\eta$.
Decreasing $ \epsilon_*/k_BT_h$ improves performance for large $c$. We conjecture that this effect is ascribed to the exponential behavior of the bath Fermi function as a function of $ \epsilon_*/k_BT$, which favors heat absorption from the hot bath as compared to heat release to the cold bath. 
The curves  are shown starting from a minimum value of 
$c$, below which  the efficiency becomes negative, indicating that the cycle no longer operates as a heat engine and requires net work input. For each $\eta$-curve the noise strength $\sigma_0$ is fixed, with $3\sigma_0\ll\beta\epsilon_*$ to ensure that the stochastic fluctuations remain small compared to the energy gap, preserving the physical operation of the thermal machine. Here we set $\sigma_0=0.01$
for $\epsilon_*/(k_B T_h)=0.1$ and $\sigma_0=0.56$
for $\epsilon_*/(k_B T_h)=5$.}
\end{figure}

\noindent We now investigate the efficiency of the engine cycle. Note that, since the density matrix remains diagonal in the energy basis throughout the cycle and the system has only two levels, $\ln\rho(t)$ is linearly dependent on $H(t)$ at all times, implying $|R_G|=1$. This ensures that $\eta_{min}=\eta=\eta_{max}$ for any $c$. This saturation is a general property of diagonal
two-level systems and does not result from optimizing
the driving protocol.
At finite $c$, the effective inverse temperature $\beta(t)$, defined by $\rho(t)\propto e^{-\beta(t)H(t)} $, differs from the bath inverse temperatures $\beta_{c}$ and $\beta_{h}$, and consequently the efficiency remains strictly below the Carnot limit. 
To account for the unavoidable uncertainty in the experimental control of the dot energy via the gate,
we introduce stochastic fluctuations in the control parameter. Specifically,
we let the energy gap randomly fluctuate as $\epsilon(t)\rightarrow \epsilon(t)+\sigma\,\xi(t)$, where $\xi(t)$ is a Gaussian white noise  with zero mean and unit standard deviation, while the noise magnitude $\sigma=k_B T \sigma_0$  is assumed to be proportional to the thermal energy through a dimensionless parameter $\sigma_0$. This stochastic contribution to the Hamiltonian \cite{10.1093/acprof:oso/9780199213900.001.0001,doi:10.1142/S1230161224500070} is an additional source of (microscopic) irreversibility and
is not associated with thermal dissipation, which is already encoded in the master equation. Yet, it does affect the transition rates $\gamma_{\uparrow,\downarrow}$ of Eq.~(\ref{mastereq}), since these depend on $\epsilon(t)$. Consequently, although within the experimental resolution the physical cycle looks just identical to the ideal one controlled by $\epsilon(t)$,  it microscopically differs from it. Only the {\it average} over the noise realizations corresponds to the physically measurable efficiency $\eta$. \\
The time interval of each transformation is discretized into $500$ equal subintervals, and both the control protocol $\epsilon(t)$ and the Gaussian noise
are modeled as piecewise constant random functions. 
The real efficiency~$\eta$ is computed by averaging over $n=100$ noise realizations, and compared with the noise-free efficiency $\eta_{max}$, which represents the ideal reference case of a full Hamiltonian control.\\

\begin{figure}[t!]
    \centering
\includegraphics[width=0.49 \textwidth]{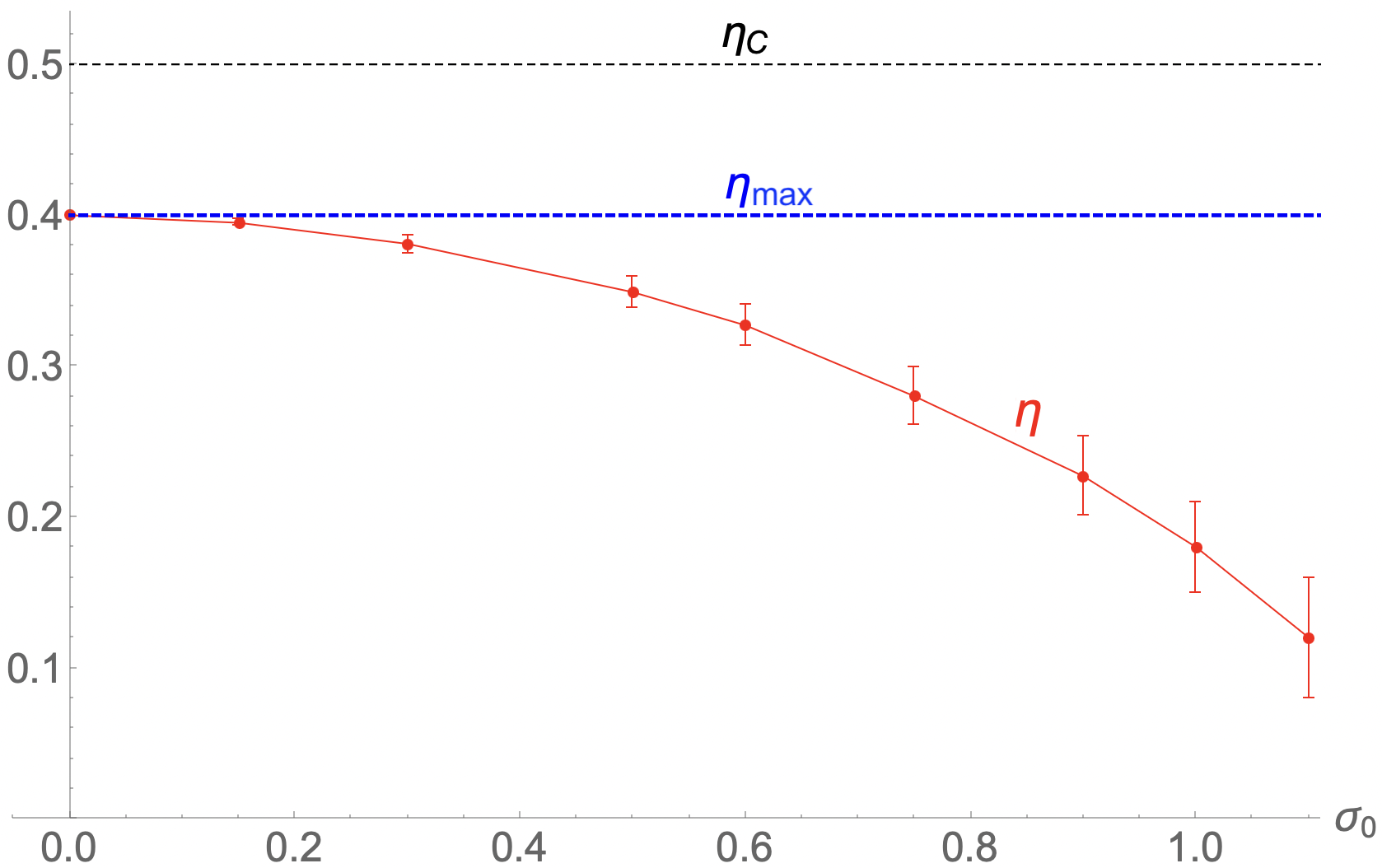}
\caption{Comparison between the efficiency $\eta$ and its bound $\eta_{max}$ as a function of the noise strength $\sigma_0$ for fixed coupling $c=30$ and $\frac{\epsilon_*}{k_BT_c}=5$ in the physically relevant regime $3\sigma_0\ll\beta_c\epsilon_*$. Error bars indicate one standard deviation of the efficiency over 100 noise realizations. In the limit $\sigma_0\to0$, $\eta$ saturates the bound, which remains below Carnot's limit. As $\sigma_0$ increases, the efficiency degrades quadratically, $\eta\simeq\eta_{max}-\alpha\sigma_0^2$, with $\alpha>0$ quantifying the sensitivity of the efficiency to control noise. This quadratic scaling is a consequence of the physical assumption of the symmetric Gaussian noise.
\label{fig:efficiencyplots}
}
\end{figure}

Figure~\ref{fig:bounds} displays the efficiencies as a function of the dimensionless coupling parameter
$c$. Carnot's efficiency is reached for an infinitely slow cycle ($c \gg 1$). Figure~\ref{fig:efficiencyplots} shows the dependence on the noise strength $\sigma_0$ at fixed $c$, compared with $\eta_{max}$ and with the Carnot efficiency $\eta_{\text{C}}$. We set $T_h/T_c=2$, yielding $\eta_{\text{C}}=1/2$.
The bound $\eta_{max}$ is saturated in the ideal, noise-free regime and remains a meaningful upper limit in the presence of stochastic energy fluctuations, thereby extending Carnot-like constraints to realistic, finite-time thermodynamic cycles.\\

\begin{figure}[t!]
    \centering
\includegraphics[width=0.48 \textwidth]{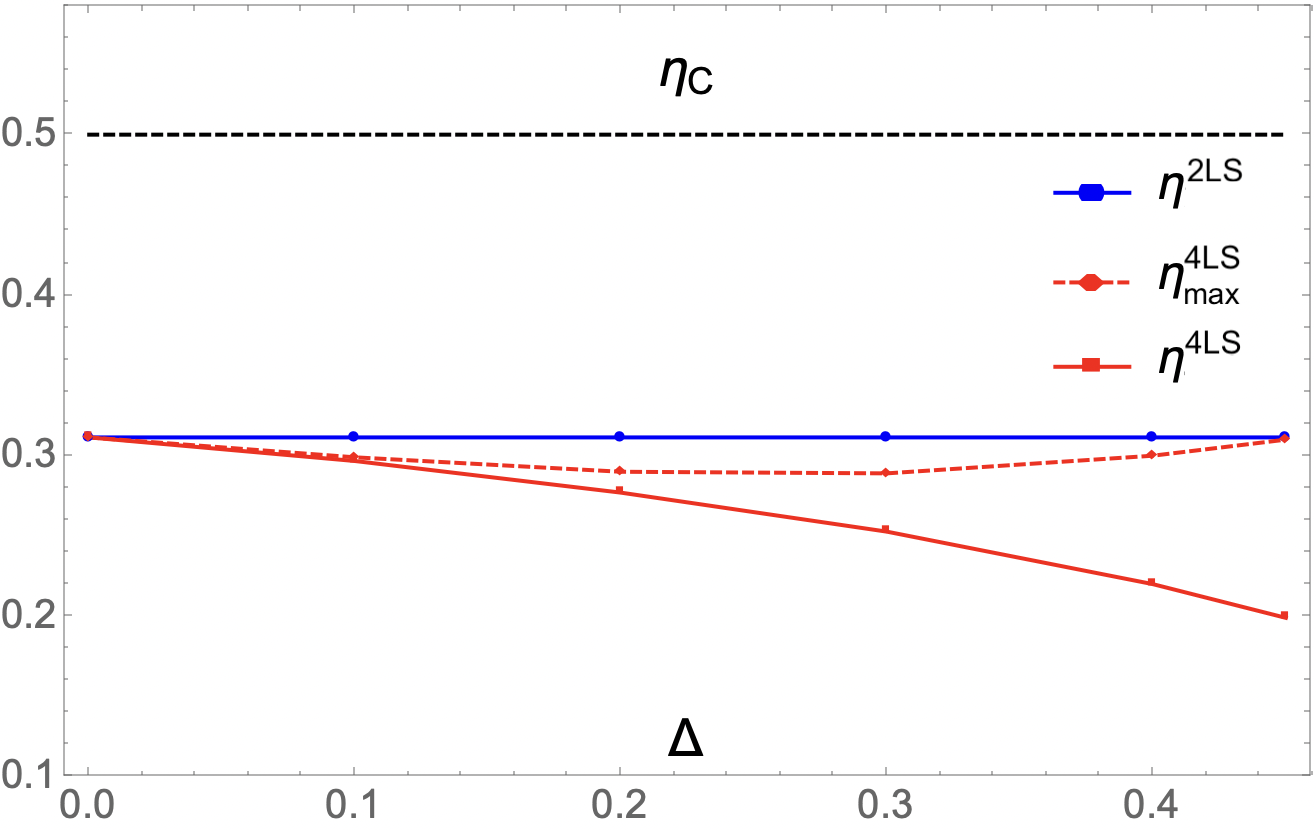}
\caption{
\label{fig:4LS} The efficiency $\eta^{4LS}$ of the four-state quantum dot and its upper bound $\eta^{4LS}_{max}$ are shown as functions of the level splitting $\Delta$, together with the single-level efficiency and the Carnot limit $\eta_C$, for a fixed coupling strength $c=15$. Increasing $\Delta$ lowers the efficiency while enlarging the gap between the efficiency and its upper bound. In the degenerate limit $\Delta\to0$, the four-state model recovers the single-level efficiency and saturates the bound $\eta^{2LS}$.}
\end{figure}

We now extend the diagonal model by introducing a
second fermionic channel. This provides an example
in which the upper and lower efficiency bounds
need not coincide.
The bath temperatures,
energy scale, and
durations are the same as in the diagonal two-level
example. The resulting system has four possible occupancy states: the empty state $(p_0)$, two singly occupied states with different energies $(p_1,p_2)$, and the doubly occupied dot state $(p_{3})$. The Hamiltonian becomes
$H_{4}(t)=
\mathrm{diag}(0,\epsilon(t),\epsilon(t)-\Delta,2\epsilon(t)-\Delta),$
where $\Delta$ denotes the energy splitting between the two electronic channels. Both channels are coupled to the same thermal baths, and the tunneling dynamics is governed by the same master equation introduced above, with two energy channels corresponding to excitation energies $\epsilon(t)$ and $\epsilon(t)-\Delta$. For a transition involving an energy $\varepsilon$, the rates are
$\gamma_{\uparrow}(\varepsilon)
=c/(1+e^{\beta\varepsilon})$ and $
\gamma_{\downarrow}(\varepsilon)
=c/(1+e^{-\beta\varepsilon}
)$. In contrast to the single-level case, the presence of two distinct energy scales $\epsilon(t)$ and $\epsilon(t)-\Delta$ generally prevents the state during the cycle from being described by an instantaneous Gibbs state with a single effective temperature. Consequently, $\ln\rho(t)$ and $H(t)$ are no longer linearly dependent, and the condition $|R_G|=1$ responsible for the simultaneous saturation of the two bounds is lost. Nevertheless, the diagonal dynamics preserves a structural constraint among the populations. Since the dynamics acts independently on the two fermionic channels, the state remains factorized throughout the evolution and the populations satisfy $p_0p_3=p_1p_2  $,
which implies 
$\text{det}(\text{corr}(V))=0$. As a consequence, the heat current saturates one
of the two local bounds. For the present dynamics, the lower
branch is followed throughout both thermal
contacts, hence the efficiency coincides with the lower
efficiency bound to numerical accuracy. As shown in Fig. \ref{fig:4LS}, increasing the level splitting decreases both the efficiency and the tightness of $\eta_{max}$, as the additional internal degree of freedom introduces deviations from the correlation structure required for saturation of the upper bound. In the degenerate limit $\Delta\to0$, the two electronic channels become equivalent, and the dynamics has  two identical single-level contributions. Moreover, the instantaneous Gibbs structure is recovered, so that $|R_G|=1$ and the two bounds collapse onto the two-level system efficiency.\\

\begin{figure}[t!]
    \centering
\includegraphics[width=0.49 \textwidth]{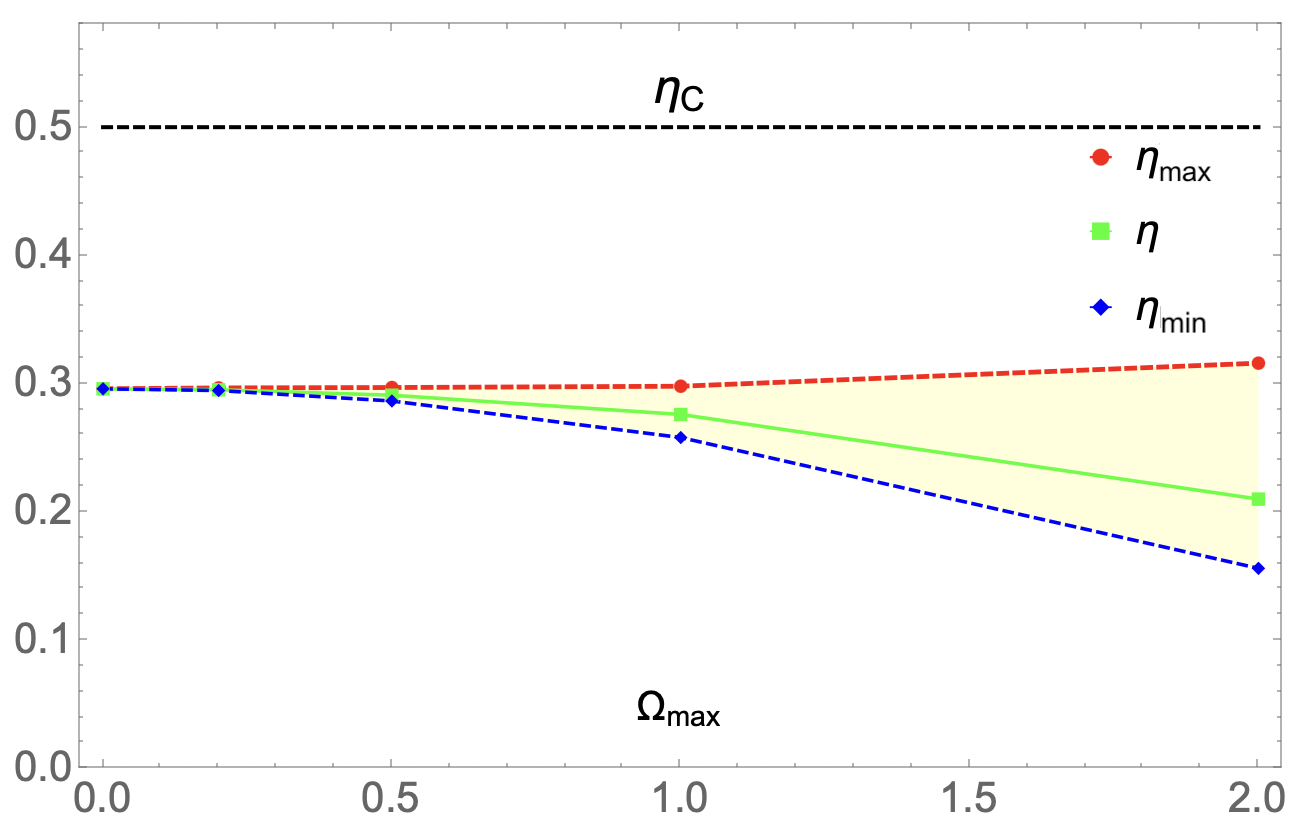}
\caption{
\label{fig:coherence} Efficiency $\eta$, upper bound $\eta_{max}$, and lower bound $\eta_{min}$ as a function of the maximum coherent driving amplitude $\Omega_{\max}$ for a fixed c=15. The width of the thermodynamically allowed region for the efficiency, depicted in yellow, reflects the gap: $\mathcal{I}_{max}(t) - \mathcal{I}_{min}(t) = 2 \,SD(H) SD(L)  \sqrt{(1 - R_S^2)(1 - R_G^2)}.$ The coherent drive progressively separates the upper and lower bounds from the actual efficiency, indicating the loss of saturation due to the additional coherence. In the classical limit $\Omega_{\max}\rightarrow0$, the two-level population model is recovered and the bound becomes saturated.}
\end{figure}

The previous examples involved purely population dynamics, with density matrices remaining diagonal in the energy basis, showing that additional internal levels modify the saturation properties of the efficiency bound. We now investigate the effects of preparing the engine in quantum superpositions of the Hamiltonian eigenbasis, by considering a two-level working medium with coherent driving. The bath temperatures,
energy scale and thermal-contact
durations are the same as in the previous
examples. The isolated transformations were treated
as instantaneous in the diagonal models, whereas
each unitary stroke in the coherent two-level
example lasts one fifth of a thermal-contact
duration.
The engine operates through a four-stroke cycle consisting of two dissipative strokes during which the system thermalizes with cold and hot baths, and two unitary strokes implementing coherent control. During the latter, the Hamiltonian includes an off-diagonal driving term, $H_{\mathrm{coh}}(t)=\frac{\epsilon(t)}{2}\sigma_z+\Omega(t)\sigma_x$,
where the coherent amplitude is modulated as $\Omega(t)=\Omega_{\max}\sin^2(\pi t/\tau_u)$ during each unitary stroke of time $\tau_u$. During the dissipative strokes, $\Omega(t)=0$ and the system evolves under the thermal Lindblad dynamics. The coherent driving generates off-diagonal density-matrix elements $\rho_{01}(t)\neq0$. These remove the population constraints present in diagonal dynamics, leading to 
$\text{det}(\text{corr}(V))\neq 0$ and preventing saturation of either efficiency bound. Increasing $\Omega_{\max}$ enhances these coherent contributions, progressively separating the efficiency from both bounds, as shown in Fig.~\ref{fig:coherence}. In the incoherent limit $\Omega_{\max}\rightarrow0$, the two-level, classical scenario is recovered, and the bound is saturated.

\section*{Outlook} We derived an information-theoretic bound on the efficiency of thermal engines. It can be tighter  than Carnot's limit $\eta_C$ in realistic, non-reversible cycles involving heat exchanges at multiple temperatures. Notably, it can be saturated by classical engines. Also, the new bound generalizes Carnot's theorem for reversible, multiple-bath  cycles.\\   
Our findings open  several routes of further investigation.
Thermal machines that saturate the bound, e.g., the two-level fermionic engine, can be  engineered with current technology \cite{exprev,dot1}.   Moreover, the bound unveils the dependence of engine efficiency on controllable properties of the working medium, including  cycle duration and energy level spacing. Hence, it is applicable to study the efficiency of thermal engines in scenarios in which Carnot's limit does not apply \cite{scully,squeeze,squeeze2}, and the performance of chemical motors \cite{chemical,motors}. Finally, its form suggests a link between engine performance and its experimental cost. In particular, an important question in the quantum regime is to determine the relation   between efficiency of heat engines  and their  {\it circuit complexity} \cite{complexity1,complexity2}, which counts the physical resources (number of gates) required to engineer quantum systems.
\section*{Acknowledgements} \label{sec:acknowledgements}
   This research was financed by the MUR-PRIN 2022 — Grant No. 2022B9P8LN-(PE3) - Project NEThEQS “Non-equilibrium coherent thermal effects in quantum systems” in PNRR Mission 4-Component 2-Investment 1.1 “Fondo per il Programma Nazionale di Ricerca e Progetti di Rilevante Interesse Nazionale (PRIN)” funded by the European Union-Next Generation EU.

\section*{Data Availability}
The data that support the findings of this article are openly available~\cite{gabetti_efficiency_data}.

\bibliographystyle{apsrev4-1}
\bibliography{references}

\end{document}